\pdfoutput=1
\documentclass[aps,pra,reprint,showpacs]{revtex4-1}
\usepackage{graphicx}
\usepackage{amsmath}
\usepackage{amssymb}
\usepackage{amsfonts}
\usepackage{dcolumn}
\usepackage{dsfont}
\usepackage{latexsym}
\usepackage{rotating}
\usepackage{color}
\usepackage{latexsym}
\usepackage{bbm}
\usepackage{subfigure}
\usepackage{float}
\usepackage{epsfig}
\usepackage{psfrag}
\usepackage{natbib}
\usepackage{bm}
\usepackage{amsthm}
\usepackage{eucal}
\usepackage{mathrsfs}
\usepackage{url}
\usepackage{braket}
\usepackage{mathtools}
\usepackage{amsmath,amssymb}
\usepackage{epsfig}
\usepackage{amsmath,amssymb}
\usepackage{hyperref}

\usepackage{color}

\usepackage{hyperref}
\hypersetup{
colorlinks=true,final=true,
        linkcolor=blue,
        citecolor=blue,
        filecolor=blue,
        urlcolor=blue,
}

\begin{document}

\title{Trajectory manipulation of an Airy pulse near zero dispersion wavelength under free carrier generated linear potential }

\author{Aritra Banerjee$^\star$ and Samudra Roy$^\dagger$}
\affiliation{Department of Physics, Indian Institute of Technology Kharagpur, W.B. 721302, India}

\email{$^\dagger$samudra.roy@phy.iitkgp.ernet.in\\
$^\star$aritra@iitkgp.ac.in}

\begin{abstract}

\noindent We investigate the dynamics of an Airy pulse that experiences free carrier generated optical linear potential in the vicinity of zero group velocity dispersion (GVD) wavelength inside a Si based-waveguide. The optically induced potential can be  realized by an inhomogeneous medium which possesses a time dependent refractive index. We propose a pump-probe scheme in Si-based waveguide where a strong continuous wave (CW) pump excites free carriers that leads to a linear potential through a time dependent refractive index change which is experienced by the  finite energy Airy pulse (FEAP) (probe). The linear potential significantly manipulate the dynamics of a FEAP and leads to a monotonous spectral shift.  We mathematically model the dynamics of the Airy pulse using linear dispersion equation containing an optical potential term and establish the general solution of the pulse for non-vanishing third order dispersion (TOD).  We derive the expression of the trajectory of FEAP which deviates significantly from its usual ballistic nature and can be tailored with the strength of the linear potential. For positive TOD, the propagating Airy pulse experiences a singularity and flips in time domain. We theoretically derive that for a specific potential strength the flipping region is squeezed to a point and revives thereafter. We propose an exact analytical solution beyond  flipping region for this specific case. Our theoretical analysis corroborates well with the numerical results.  The present study may be useful in applications related to pulse reshaping and trajectory manipulation.

\end{abstract}

\maketitle

 \section{Introduction}

\noindent   The infinite energy Airy wave packet was first introduced as a solution of the Schr\"{o}dingers equation in free space in 1979 \cite{Berry}. Almost thirty years later the concept of finite energy airy beam (FEAB) which contains a decay factor was introduced in the context of optics \cite{Siviloglou} and it was also experimentally realised \cite{Siviloglou_b}. Since this remarkable discovery many works have been reported exploiting the  unique features of FEAB like self acceleration, quasi diffraction free and self healing nature \cite{Siviloglou,Siviloglou_b,Broky}. The properties of FEAB in nonlinear regime is also explored extensively where beam reshaping occurs  \cite{Rui,Lotti,Lotti_b}. One is to one correspondence between the spatial diffraction and the temporal dispersion, leads to the concept of finite energy Airy pulse [FEAP], which is introduced recently as a temporal analogous of Airy beam \cite{Saari}. Airy pulses are the only waveforms which maintain its width and amplitude in a linear dispersive medium and robust against perturbation. After the discovery of the self healing property of the Airy pulse, several interesting works have been done in the temporal domain like supercontinuum generation \cite {Ament}, absolute focusing under third order dispersion (TOD) \cite {driben,Shaarawi}, mimicking event horizon through Airy-soliton collision \cite {Yang} etc. The natural trajectory of the Airy pulse is parabolic as the temporal position of the main lobe of a FEAP, is a quadratic function of the space it sweeps.  This is strictly true when FEAP experiences only second order group velocity dispersion (GVD). The regular trajectory of the moving pulse is influenced significantly under higher order dispersions and it can even   be controlled by suitable dispersion engineering \cite{Bai}. However efficient dispersion engineering is a challenge. For  the last few years several works have been reported which discussed about the manipulation of the trajectory of the beam or pulse without hampering the waveguide parameters \cite{Liu,Yiqi,Nikolas,Han}. The introduction of linear optical potential is found to be an unique technique to control the dynamics of an Airy beam \cite{Liu,Yiqi}. Using the similar concept, the role of the optical potentials on the dynamics of the Airy pulse is investigated in temporal domain where cross phase modulation (XPM) is used to create a potential that varies linearly or quadratically with time \cite{Han}. The realization of the optical temporal potentials opens up new avenues in the study of the FEAP dynamics. In the previous study \cite{Han} the effect of higher order dispersions on the pulse dynamics is neglected. It was considered that, the duration of the pulse is long, hence the effects of the higher order dispersions are insignificant. However, this assumption doesn't hold good specially when the pulse experiences a time dependent optical potential. The linear potential leads to a unidirectional frequency shift during propagation and because of that the pulse should experience higher order dispersion.  It is obvious that, if the pulse is launched near to the zero group velocity dispersion (GVD) point, after moving a finite distance, it will face strong third order dispersion owing to its unidirectional frequency shifting. Hence it is important to address this issue in order to investigate the full dynamics of the FEAP under linear potential.
   
  In this report, we propose a simple pump-probe scheme inside a Si-waveguide where a FEAP (probe) experiences dynamic refractive index change created through strong CW wave (pump).  The CW wave induces free carriers (FCs) through the process of two photon absorption (TPA) \cite{Gaafar}. The optically generated FCs lead to a dynamic refractive index change which may act as an optical potential for the FEAP. We critically design the geometry of a Si-based waveguide so that it exhibits two zero GVD profile. The dispersion slope takes alternative signs at zero dispersion points that leads to positive or negative TOD coefficients. It is well-known that under positive third order chromatic dispersion the Airy pulse faces a singularity and the temporal distribution of the pulse reverses after a certain distance \cite{driben}. On the other hand if the chromatic dispersion is negative, the robustness of the Airy pulse is enhanced for a larger distance. In this report we study both the cases of positive and negative third order dispersions when the pulse is subjected to the linear potential.  Introduction of the potential significantly influence the dynamics of the FEAP. An extra degree of freedom is achieved through the strength of the potential that can be used to engineer the trajectory of the pulse. We derive a full analytical solution of a FEAP that moves under linear potential and theoretically investigate how the pulse trajectory is controlled by the potential strength. Further we establish the general solution of the FEAP for negative third order dispersion and derive the analytical expression of the pulse trajectory that agrees well with the numerical simulations. We extend our theory for positive TOD which is interesting since the Airy pulse experience a singularity at certain propagation distance. By dividing the entire dynamics of the Airy pulse into three regions we derive respective analytical solutions that describe the complicated dynamics of the Airy pulse. The analytical solutions derived by us, facilitate in understanding the peculiar dynamics of the Airy pulse under linear potential for positive dispersion where the pulse flips and moves with a reverse acceleration. All out theoretical expressions are verified by solving the pulse propagating equation numerically.

   \section{Mathematical model}
\begin{figure}[h!]
 \begin{center}
  \includegraphics[trim=0.0in 0.0in 0.0in 0.0in,clip=true,  width=70mm]{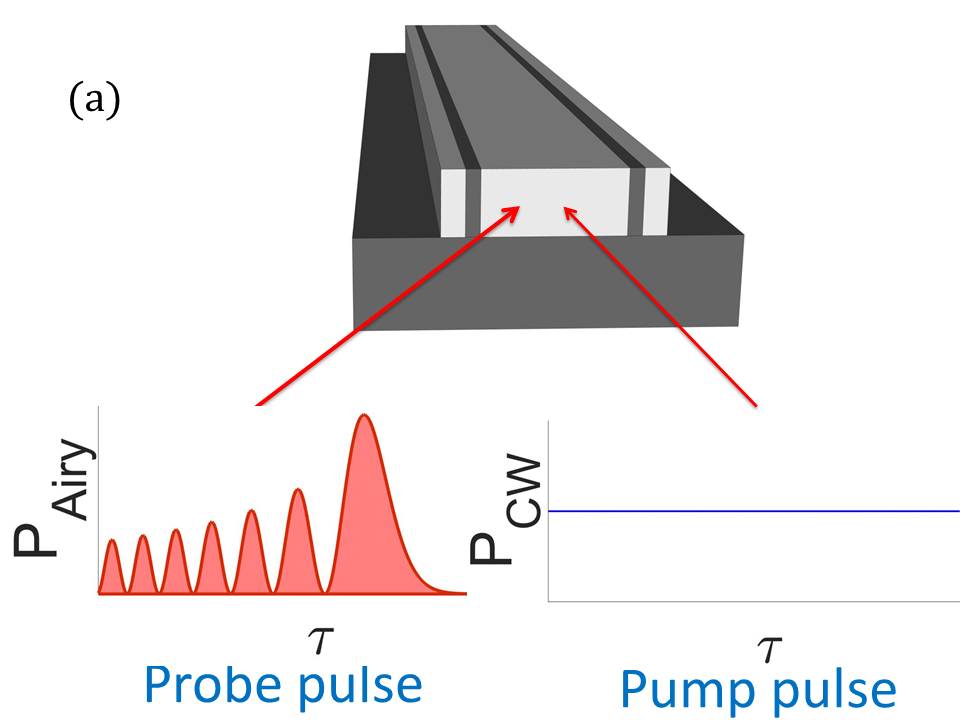}
  \includegraphics[trim=0.0in 1.6in 0.0in 0.0in,clip=true,  width=70mm]{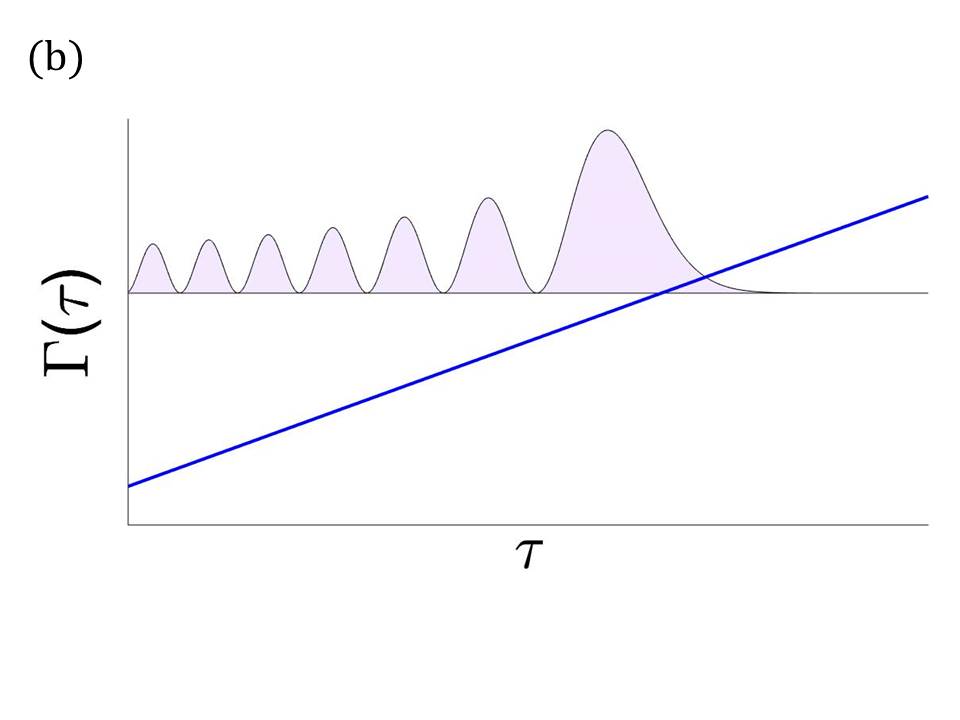}
  \vspace{-1em}
 \caption{(a) Schematic diagram of a pump-probe system in a Si-waveguide. (b) The linear potential ($\Gamma$)  experienced by the FEAP through free-carrier induced dynamic refractive index change $\delta n(t)$ in normalised units.    }
                 
\label{Figure1}
\end{center}
 \end{figure}

\noindent An optical potential can be realised by the temporal variation of refractive index \cite{Fabio}. The temporal variation of the refractive index influences the propagation constant of the pulse which in return affects the dynamics of the pulse. In a recent work, the phenomenon of temporal reflection and refraction is explained, where reflection takes place at the temporal boundary across which refractive index changes \cite{Plansinis}. In order to realize the optically induced time dependent potential, we propose a scheme where a strong CW pump ($\psi_0$) is injected inside a Si-based waveguide along with a FEAP (see Fig.\ref{Figure1}(a)). The CW pump induces free carriers (FCs) through the rate equation \cite{Agarwal}:  
 
\begin{equation} \label{eq1}
     \frac{\partial N_c}{\partial t} =\frac{\beta_{TPA}}{2h\nu_0}|\psi_0|^4-\frac{N_c}{t_c} ,
     \end{equation}
     
where $N_c$, $\beta_{TPA}$ and $t_c$ are free-carrier density, TPA coefficient and recombination time, respectively. Normally, $t_c \sim 1$ ns  is very large compared to the pulse width which is typically $\sim 100$ fs. Hence, neglecting the carrier recombination time we can have the temporal dependence of carrier density as, $N_c(t)=\rho_{TPA}|\psi_0|^4t$, where, $\rho_{TPA}=\beta_{TPA}/2h\nu_0$. The time dependent carrier density leads to a refractive index change as $\delta n(t)=k_c N_c=\Delta n_t  t$, where $k_c=1.35 \times 10^{-21}$ cm$^3$ \cite{Lipson} and $\Delta n_t = \rho_{TPA} k_c|\psi_0|^4$.

\noindent In our study we consider FEAP is propagating in a dispersive medium with dispersion relation $\beta(\omega)$. The temporal variation of refractive index with the form $n=n(\omega)+\Delta n_t t$  modifies the dispersion relation as, $\tilde{\beta}(\omega)=\beta_0+\beta_1(\omega-\omega_0)+(\beta_2/2!)(\omega-\omega_0)^2+(\beta_3/3!)(\omega-\omega_0)^3+\beta_\mu t$, where $\beta_\mu=k_0\Delta n_t$. $\Delta n_t (= \rho_{TPA} k_c|\psi_0|^4)$ is the index change per unit time and $k_0=\omega_0/c$. In normalised units, the dynamics of FEAP  with amplitude $U(\xi,\tau)$ near zero-GVD point under a linear potential $\Gamma(\tau)=f\tau$ can be modelled as,

   \begin{equation} \label{eq2}
     i\frac{\partial U}{\partial \xi} +\sum_{n\geq2}^{3} i^n \delta_n \frac{\partial^n U}{\partial \tau^n} +\Gamma(\tau) U=0   
     \end{equation} 
     
\noindent The actual amplitude of the envelope is rescaled as  $u=\sqrt{P_0}U$. $z$ and $t(=T-zv_g^{-1})$ are the space and time variables in the frame that is moving along the pulse with a group velocity $v_g$ where $T$ is the time in the rest frame.   The other parameters are normalized as,  $\xi=z{L_D}^{-1}$, $\tau=t/t_0$. $P_0$, $t_0$ and $L_D=t_0^2|\beta_2|^{-1}$ are the input peak power, width of the primary lobe of the initial FEAP and dispersion length respectively.  $\delta_2$ is related to the numeric sign of GVD parameter as, $\delta_2=sgn(\beta_2)/2$. The TOD parameter ${\beta_3}$ is rescaled as, $\delta_3=\beta_3/(3!|\beta_2|t_0)$. The linear potential coefficient is normalised as, $f=\beta_\mu t_0L_D$, where $\beta_\mu$ is the unnormalized potential coefficient. In Fig.\ref{Figure1}(b) we schematically show how an FEAP experiences a optical linear potential generated due to the free-carrier induced dynamic refractive-index change.  The solution of  Eq.\eqref{eq2}  without the perturbation of third order dispersion (diffraction) was discussed in a recent study in the context of Airy beam \cite{Liu}.  We consider the initial pulse is of the form $U\left( 0,\tau  \right)=Ai(\tau)\exp(a\tau)$, where $a$ is the truncation parameter which makes the infinite energy pulse a finite one. It can be shown that in absence of $\delta_3$, the pulse evolves as,

     \begin{equation}\label{eq3}
     \begin{aligned}
     U\left( \xi ,\tau  \right)=~Ai\left[ \tau -\left( 1+2f \right)\frac{{{\xi }^{2}}}{4}+ia\xi ~ \right] \times \\
     \exp \left[ a\left( \tau -\frac{\xi^2}{2} (f+1)+\frac{a^2}{3} \right) \right]~\text{exp}\left( i\Theta  \right)
     \end{aligned}
     \end{equation}

      \begin{figure}[h!]
      \begin{center}
      \includegraphics[trim=1.3in 0.0in 1.2in 0.0in,clip=true,  width=100mm]{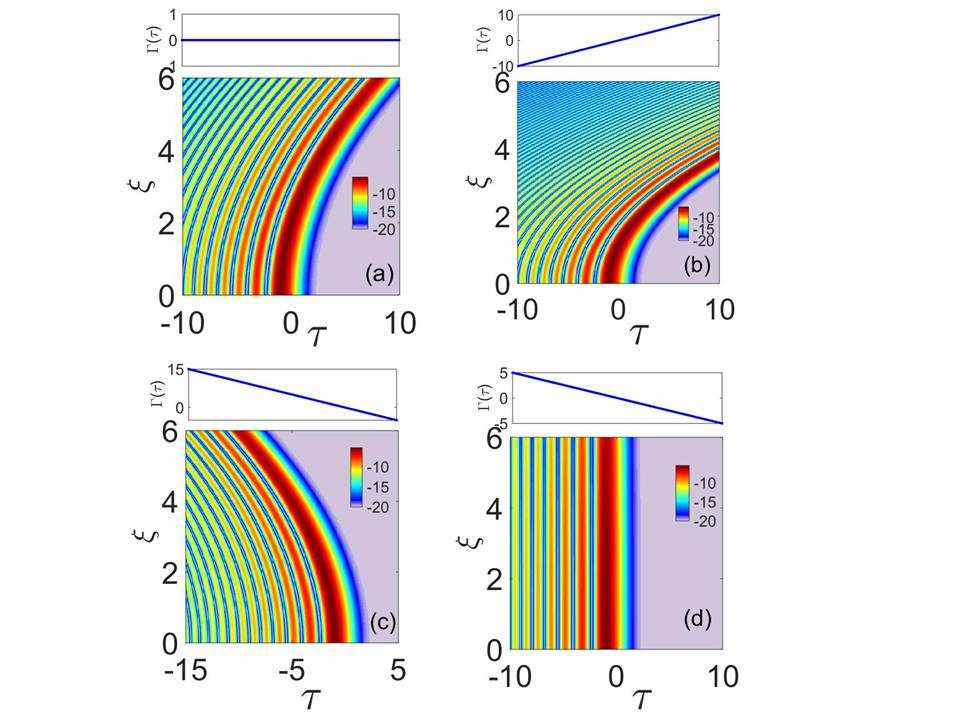}
      \vspace{0em}
      \caption{Temporal trajectory of the pulse with $\delta_3=0$ for different values of potential strength $(f)$ : $(a)f=0,(b)f=1,(c)f=-1,(d)f=-0.5$ . The variation of the potential $\Gamma(\tau)$ is given on the upper panels of the figure.} 
      \label{Figure2}
      \end{center}
      \end{figure}
\noindent where, $\Theta=[{a^2\frac{\xi}{2}+(f-\frac{1}{2})\tau \xi}- (1+3f+2f^2)\frac{\xi^3}{12}]$  denotes the phase of the propagating pulse. It is obvious that, in absence of any perturbations (here, $\delta_3, \Gamma$) the Airy pulse exhibits self-acceleration in $\tau-\xi$ plane along a parabolic trajectory $\tau_p \approx \tau_{p0}+\xi^2/4$, where $\tau_p$ is the temporal position of the main lobe of the propagating Airy pulse and $\tau_{p0}\approx -(3\pi/8)^{2/3}$ denotes the initial temporal position of the main lobe . However, it is evident that, the trajectory of the propagating Airy pulse is affected in presence of linear potential and  $\tau_p$ becomes,    $\tau_p \approx \tau_{p0}+(1+2f)\frac{\xi^2}{4}$. The expression of the temporal position clearly indicates that, depending on the value of the potential strength $f$ one can manipulate the trajectory of an Airy pulse. The acceleration of the Airy pulse can also be altered by changing the numeric sign of $f$. In Fig.\ref{Figure2} we  plot the  temporal dynamics of FEAP subjected to different values of $f$. The manipulation of the trajectory is evident for different values of $f$ in $\tau-\xi$ plane. It is interesting to note, the  acceleration of the Airy pulse in $\tau-\xi$ plane which is,  $\frac{d^2\tau}{d\xi^2}=f+1/2$  can be nullified  for $f=-1/2$  and therefore the pulse moves in a straight line trajectory (see Fig.\ref{Figure2}(d)).     
  
 \section{Dynamics under Third order dispersion} 
   
\begin{figure}[h!]
                 \begin{center}
                 \includegraphics[trim=0.8in 0.1in 1.0in 0.1in,clip=true,  width=93mm]{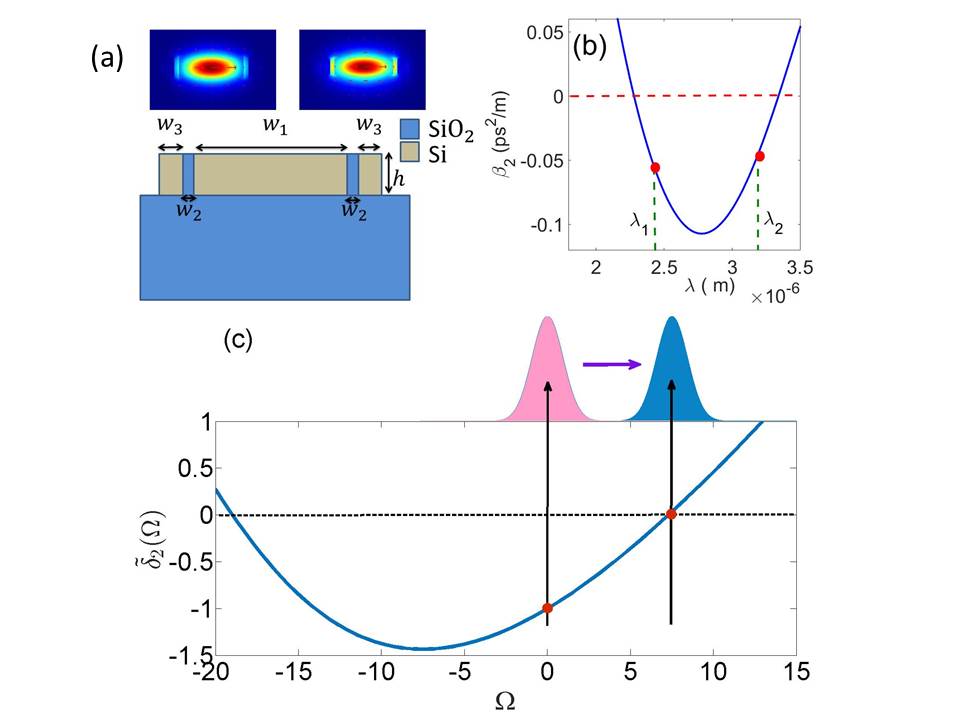}
               \vspace{-1em}  
    \caption{(a) The block diagram shows the geometry of the proposed Si-based waveguide. The dimensions are $w_1$ = 1500 nm; $w_2$ = 110 nm; $w_3$ = 230 nm  and $h$ = 410 nm. The confinement of the fundamental TE mode at the operating wavelengths ($\lambda_1=$ 2425 nm and $\lambda_2=$3200 nm) are shown in the inset. (b) Dispersion profile of the proposed waveguide where the operating wavelengths are indicated by the red solid dots. At operating wavelength $\lambda_1$, the TOD parameter is positive and at $\lambda_2$  TOD parameter is negative. (c) The variation of normalised GVD parameter ($\tilde{\delta}_2$) with normalised frequency ($\Omega=(\omega-\omega_0)t_0$) where it is shown how the input spectra (at $\Omega=0$) is blue shifted under positive linear temporal potential during propagation. }
                 
     \label{Figure3}
\end{center}
 \end{figure}

\noindent The power-law optical potential is useful in manipulating the trajectory of FEAP \cite{Han}. A unidirectional frequency shift takes place when the FEAP moves under linear potential. It is important to note that, the investigation of the role of linear potential on pulse dynamics is incomplete if we ignore the relevance of TOD, specially when the pulse experiences a frequency shift. In order to realize the importance of TOD in Fig \ref{Figure3}(c) we demonstrate the spectral shift of an FEAP (under a linear potential) against normalised GVD $\tilde{\delta}_2(\Omega)$. The initial spectra of the input FEAP centred at $\Omega=0$ moves towards high frequency side as it propagates inside the waveguide. From this illustration it is evident that, the moving Airy pulse experiences a strong TOD as a consequence of spectral shift under linear-potential. For negative potential strength the spectra will move at the low frequency side and encounter the first zero GVD point.  Hence for realistic study on linear potential, it is essential to include TOD effect in the pulse dynamics. The temporal dynamics of the FEAP changes radically near the zero-GVD wavelength where TOD is not negligible \cite{driben}. The manipulation of the trajectory of the FEAP using linear potential becomes more effective when $\delta_3 \neq 0$.  The slope of the GVD curve at particular wavelength determines the sign of the TOD parameter ($\delta_3$). For positive TOD the Airy pulse faces a singularity after propagating a certain distance  and   propagates with the flipped temporal tail \cite{driben}. On the contrary the pulse becomes robust and propagates a longer distance without distortion in case of negative TOD \cite{Shaarawi}. We study both the cases individually in the context of linear potential. We propose a Silicon based waveguide which shows two zero GVD profile. The two-zero GVD profile is advantageous in order to study the effect of positive and negative TOD together in a single waveguide. The waveguide geometry and related dispersion profile are shown in Fig.\ref{Figure3}(a) and Fig.\ref{Figure3}(b) respectively. The slope of the dispersion curve as illustrated in Fig.\ref{Figure3}(b) is altered across the dispersion minima. So, $\beta_3$ values with opposite numeric signs at two different launching wavelengths can be achieved easily for given GVD profile (as indicated by the red dots in Fig.\ref{Figure3}(b)). 

\noindent We have evaluated the values of GVD parameter ($\beta_2$) and TOD parameter ($\beta_3$) in real units by using COMSOL multi-physics software for the proposed waveguide. The values of GVD and TOD coefficients are $\beta_2$=-0.038  ps$^2$/m and $\beta_3$=0.0011 ps$^3$/m respectively at the first launching wavelength $\lambda_1$= 2425 nm. For the second launching wavelength, $\lambda_2$=3200 nm, the values are $\beta_2$=-0.043  ps$^2$/m and $\beta_3$=-0.0015 ps$^3$/m. Note that,the numeric signs of $\beta_3$ are opposite for two launching wavelengths which is desirable in our study. The values of the normalised TOD parameter that we have used in our simulation ranges  between 0.05 to 0.1 which can be achieved for the pulse-width ranging between 50 fs to 100 fs. (e.g. for $\lambda_1$, $\delta_3 \approx$ 0.08  for $t_0 =$ 60 fs ; similarly, for $\lambda_2$, $\delta_3 \approx$ -0.08  for $t_0 =$ 75 fs).

 \subsection{Negative TOD  ($\delta_3<0$)}   
   
 \noindent The FEAP ($a \neq 0$) is not the exact solution of the dispersive equation with linear potential term. However it is possible to derive an analytical expression of the truncated Airy pulse moving under TOD and linear potential as, 

   \begin{figure}[h!]
   \begin{center}
   \epsfig{file=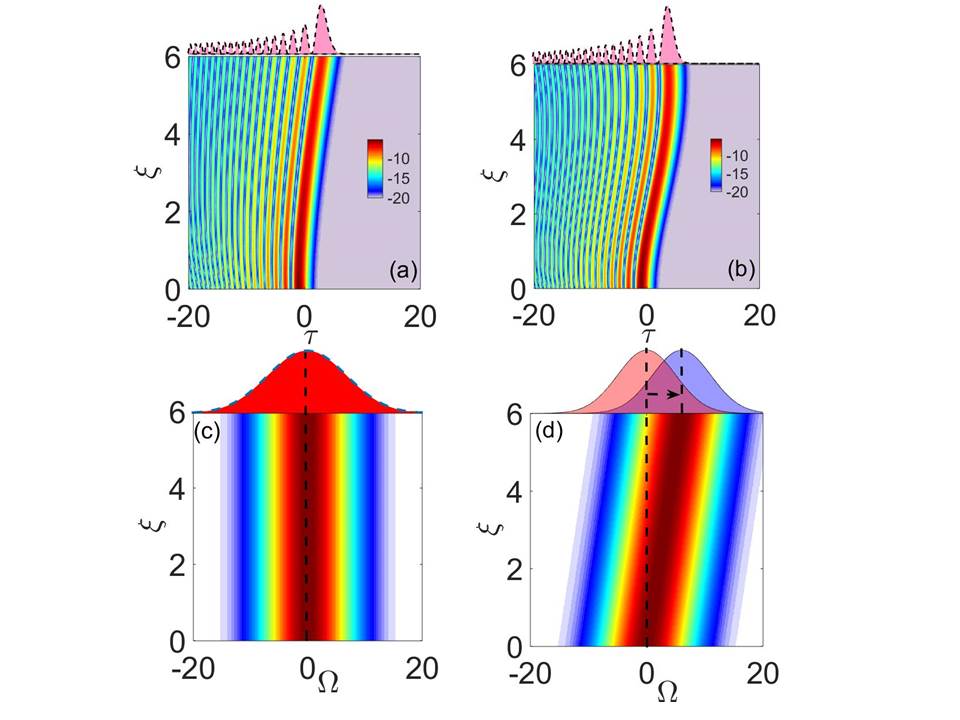,trim=1.35in 0.0in 1.55in 0.0in,clip=true, width=90mm}
  \vspace{-1em}
  
   \caption{ Density plot of the propagating FEAP based on the numerical solution of 
   Eq.\ref{eq2} for (a)$f=0$ and (b) $f=+1$ . We compare the analytical solution Eq.\ref{eq4}  and numerical output (dashed line) in the upper panel. The shift in frequency in presence of $f$ is compared in (c)$f=0$ and (d)$f=1$. The frequency shift is entirely dependent on $f$ and independent of TOD.}  
   \label{Figure4} 
   \end{center}
   \end{figure}
  
\begin{equation}\label{eq4}
     U(\xi,\tau)=\frac{1}{c}\exp\left(\frac{a^3}{3}\right)Ai\left(\frac{b}{c}-\frac{n^2}{c^4}\right)\exp i\left(\frac{2n^3}{3c^6}-\frac{nb}{c^3}+\phi \right), 
     \end{equation}

\noindent where, $n=ia-\frac{\xi}{2}-\frac{3}{2}{{\delta }_{3}}f{{\xi }^{2}}$, $b=\tau -{{a}^{2}}-\frac{f\xi^2}{2}-{{\delta }_{3}}{{f}^{2}}{{\xi }^{3}}$ and $c=(1+3|\delta_3|\xi)^\frac{1}{3}$. The phase of the propagating pulse is,  $\phi (\xi,\tau) =\phi_0(\xi)+f\xi\tau$, where  $\phi_0 (\xi)=-\frac{1}{6}f^2 {{\xi }^{3}}-\frac{1}{4}\delta_3 f^3\xi^4$. The analytical solution derived in Eq.\ref{eq4} is verified by direct numerical simulation for a fixed TOD parameter and they agree well (upper panels of Fig .\ref{Figure4}(a) -(b)). From the phase information of the analytical solution we find that the presence of linear potential results a linear frequency shift $\Delta \Omega=f\xi$, which is illustrated in Fig. \ref{Figure4}(c) and (d).  The potential term present in the argument of the propagating Airy pulse modifies the trajectory significantly. In Fig.\ref{Figure5} we plot the temporal dynamics of the pulse for different  potential strengths ($f$) with $\delta_3<0$. In Fig.\ref{Figure5}(a) the pulse does not experience any optical potential and it follows a nearly parabolic trajectory which is expected. But the trajectory deviates from the parabola in Fig.\ref{Figure5}(b)-(d) when $f$ is non-zero. Hence,we have an additional parameter in terms of potential strength $f$ that can manipulate the trajectory of the pulse.  The general form of the temporal position of the primary lobe of the FEAP is evaluated as,
  \begin{equation}\label{eq5}
  \tau_p=\tau_{0p}+\frac{1}{2}f\xi^2+\delta_3f^2\xi^3+\frac{\Delta}{c^3}
  \end{equation}

   \begin{figure}[h!]
    \begin{center}
    \epsfig{file=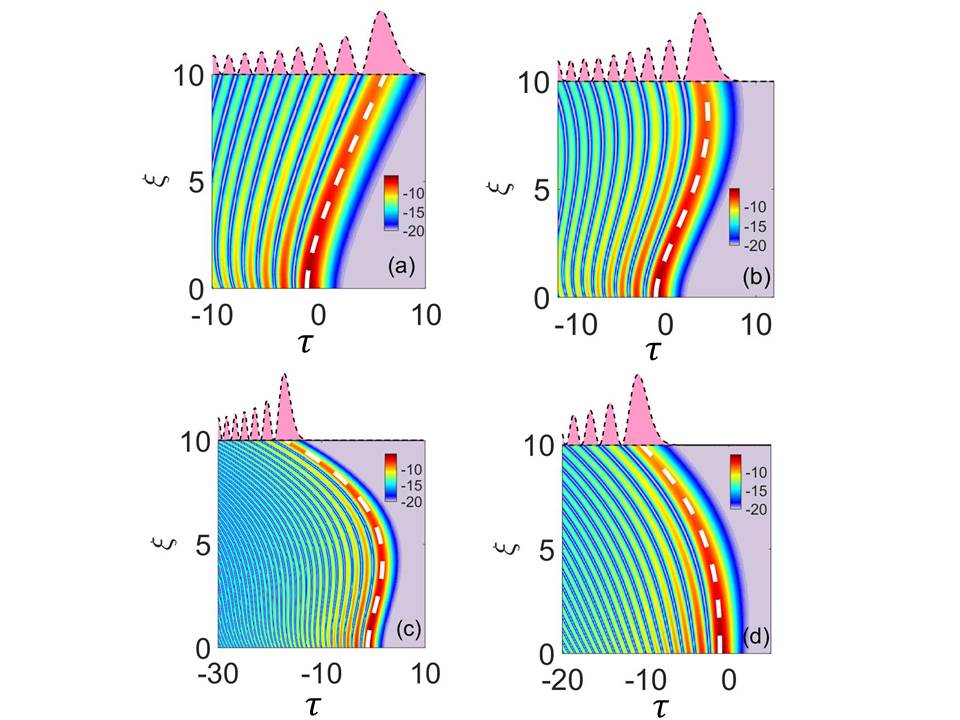,trim=1.5in 0.0in 1.7in 0.1in,clip=true, width=90mm}

   \vspace{0em}
   \caption{The propagation dynamics of FEAP for different potential strength ($(a)f=0,(b)f=0.5,(c)f=1.0 (d)f=-0.5$) at a fixed value of TOD parameter($\delta_3=-0.08$).The white dashed lines in the density plots represent the trajectory of the primary lobe of the pulse. The numerical solutions at the output (black dashed lines) is verified with the analytical results (pink shaded figure) in each cases.} 
   \label{Figure5}
   \end{center}
   \end{figure}
\begin{figure}[h!]
       \begin{center}
       \epsfig{file=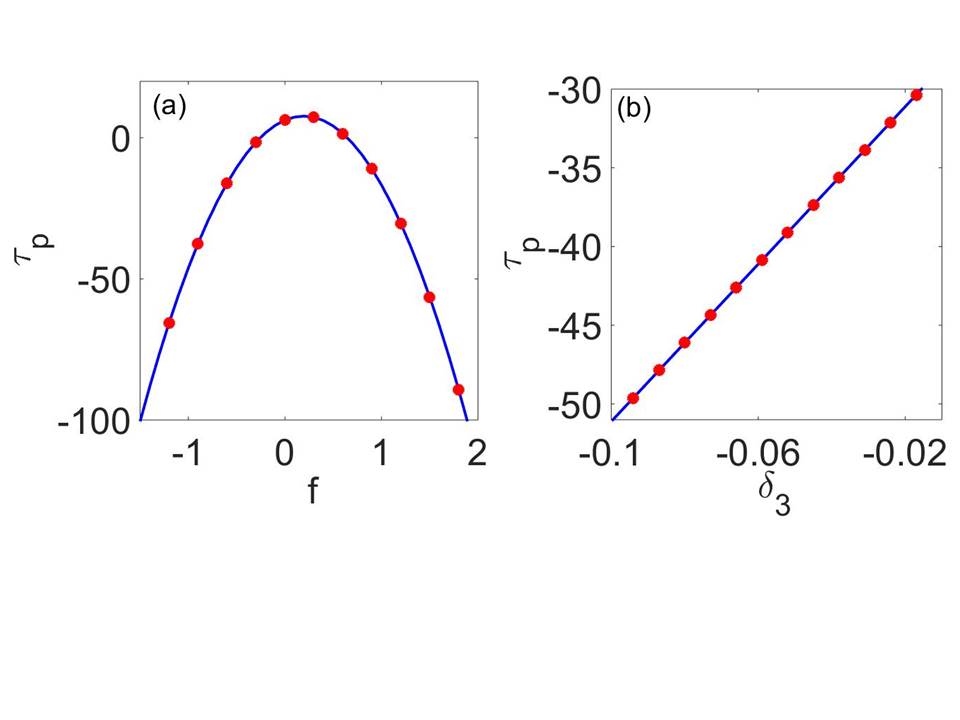,trim=0.1in 2.1in 0.1in 0.7in,clip=true, width=88mm}
       
     \vspace{0em}
      \caption{(a) The variation of primary lobe position ($\tau_p$)with the strength of the potential ($f$) for $\delta_3=-0.08$ at fixed distance $\xi=10$.(b) The variation of primary lobe position ($\tau_p$)with $\delta_3$ for $f=-1$ at fixed distance $\xi=10$.   } 
      \label{Figure6}
      \end{center}
      \end{figure}


 \noindent where  $\Delta=\frac{\xi^2}{4}+\frac{3}{2}\delta_3f\xi^3+\frac{9}{4}\delta_3^2f^2\xi^4$.  Eq.\ref{eq5} provides us the insight about the temporal trajectory of the FEAP under TOD and linear potential. It is evident that the natural quadratic dependence of $\xi$ is no longer valid for non-zero $f$. The new power law trajectory is sensitive to the numeric sign of $f$ as well as $\delta_3$. We illustrate the numerically obtained temporal dynamics of FEAP  for different  potential strength in Fig.\ref{Figure5} for a fixed $\delta_3$. Based on the solution derived in Eq.\ref{eq4}, we theoretically calculate the trajectory (Eq.\ref{eq5}) of the moving pulse (white dashed lines in the mesh plots of Fig.\ref{Figure5}) and it agrees well with the numerical path. The analytical envelopes (black dashed line in the upper panel) are also in agreement with the numerical output (shaded region). In Fig. \ref{Figure6}(a) and (b) we plot the variation of $\tau_p$ as a function of potential strength $f$ and TOD parameter $\delta_3$, respectively at a fixed distance. The theoretical expression of $\tau_p$ as denoted by the solid lines in \ref{Figure6}(a) and (b) agrees nicely with the numerical results (solid dots). The temporal location of the main lobe of the FEAP at output largely depends on $f$. However we can define a critical $f =1/|{\delta_3}|\xi(3+{c^3})$ for which the deviation of the FEAP is minimal. Note that, for fix $\xi$,  $\tau_p$ changes linearly with $\delta_3$ (see Fig. \ref{Figure6}(b)). These might be some useful information in manipulating the trajectory of the FEAP.  
 
\subsection{Positive TOD ($\delta_3>0$)}

\begin{figure}[h!]
    \begin{center}
    \epsfig{file=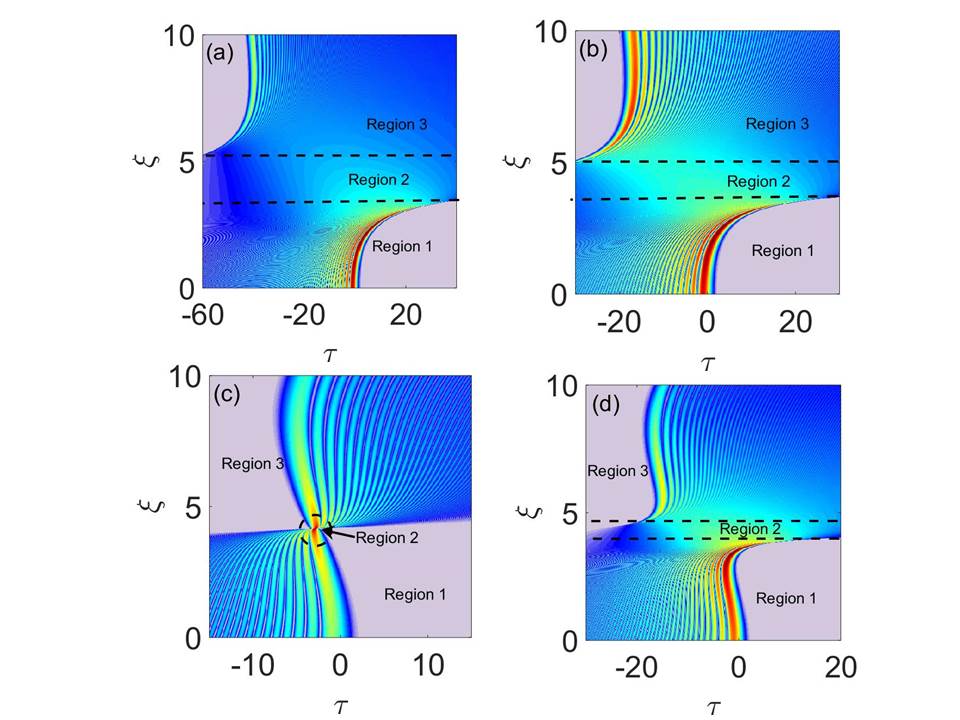,trim=1.2in 0in 1.0in 0.0in,clip=true, width=88mm}
    \vspace{0em}
   \caption{The dynamics of the pulse for (a)$f=0.5$ (b)$f=0$ (c)$f=-1$ and (d)$f=-1.5$ with a fixed positive TOD coefficient $\delta_3=0.08$. Three distinct regions are marked in the figure. The region between the marked line increases  for $f=0.5$ and decreases for $f=-1.5$ and it squeezes to a point for $f=-1$ .        } 
   \label{Figure7}
   \end{center}
   \end{figure}

\noindent In this section we investigate the role of linear potential on Airy dynamics considering the slope of GVD as positive ($\delta_3>0$). For $\delta_3>0$ , the parameter $c$ becomes $c=(1-3\delta_3\xi)^\frac{1}{3}$, which means Eq.\ref{eq4} will experience  a singularity at some specific distance $\xi_{flip}=(3\delta_3)^{-1}$. The natural dynamics of the Airy pulse is interrupted by this singularity and it experiences a temporal flipping followed by a reverse acceleration \cite{driben}.  In  Fig.\ref{Figure7} we illustrate the temporal dynamics of an Airy pulse for $\delta_3>0 (=0.08)$ with different potential strength. The over all dynamics  is divided into three regions, (i) Region 1 : initial ballistic trajectory, (ii) Region 2: the singularity zone and (iii) Region 3: the temporal flipping. The potential strength $f$ doesn't change the spatial position of the flipping point $\xi_{flip} (\approx 4.16)$ but influences Region-2 by reshaping the area of the singularity zone. Region 2 can be expanded (Fig.\ref{Figure7}(a)) or compressed (Fig.\ref{Figure7}(d)) depending on the value of $f$. Even the area can be compressed to a single point for a specific value of $f$(fig.\ref{Figure7}(c)) . In the following sections we describe in detail the behaviour and trajectory of the pulse separately for all three regions.

  \subsubsection{Region 1 ($\xi<\xi_{flip}$)}

\begin{figure}[h!]
    \begin{center}
    \epsfig{file=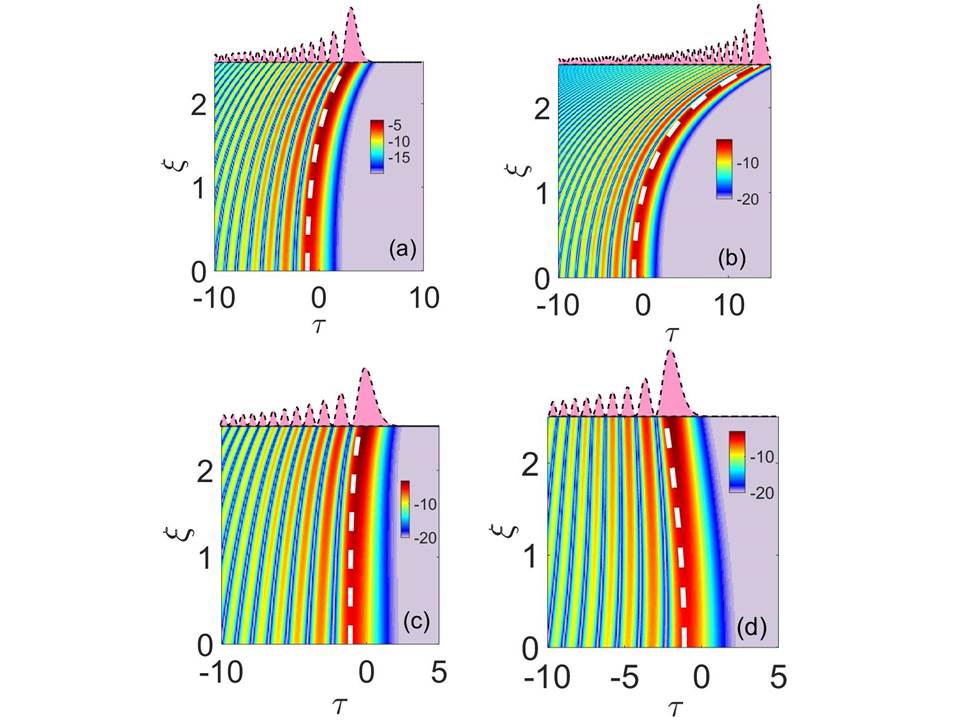,trim=1.7in 0.0in 1.5in 0.0in,clip=true, width=90mm}
   \vspace{0em}
   \caption{The temporal dynamics of the pulse  for  (a)$f=0$ (b)$f=0.5$ (c)$f=-0.5$ and $f=-1$ with fixed $\delta_3=0.08$.The white dashed lines represent the trajectories of the primary lobe of the pulse for different $f$. The numerical form of the solution of the output pules (black dashed lines) is verified with analytically obtained results and is shown on the upper panels of the plots.    } 
   \label{Figure8}
   \end{center}
   \end{figure}

\begin{figure}[h!]
    \begin{center}
\epsfig{file=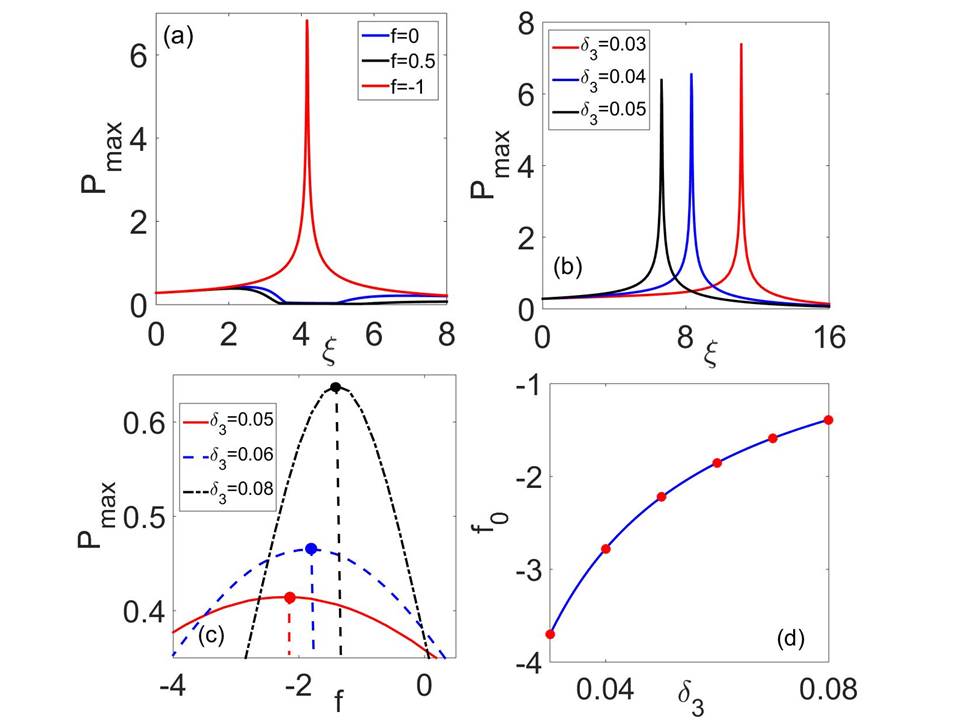,trim=0.8in 0.0in 0.5in 0.0in,clip=true, width=90mm}

   \vspace{-1em}
  \caption{(a)The variation of maximum power ($P_{max}$) with $\xi$ for different potential strength with fixed $\delta_3=0.08$. (b) The variation of $P_{max}$ with $\xi$ for different sets of $\delta_3$ with fixed $f=-1$.(c) The variation of $P_{max}$ with $f$ at a fixed distance $\xi=3$ with changing $\delta_3$. The colored dashed lines indicate the critical values ($f_0$) of $f$ for which the power is maximum. (d) The variation of the  $f_0$ with $\delta_3$ for a fixed $L (=3)$.         } 
  \label{Figure9}
   \end{center}
   \end{figure}

Before facing the singularity (i.e $\xi<\xi_{flip}$) the analytical form of the propagating FEAP is same as Eq.\ref{eq4} with $c=(1-3\delta_3\xi)^\frac{1}{3}$.
 The general solution reveals that, the trajectory of the pulse  can  be deviated from the usual ballistic trajectory $(f=0)$ if the strength of the potential ($f$) is introduced. It is in fact quite similar to the case of the negative TOD. In Fig.\ref{Figure8} we show the trajectory of the Airy pulse for different values of $f$ for a fixed $\delta_3 (=0.08)$ in the regime $\xi<\xi_{flip}$. The acceleration of the pulse is efficiently controlled by the strength of the potential.  The temporal location of the main lobe is mapped analytically (white dashed line) by using Eq.\ref{eq5} which agreers well with  simulated trajectory indicated by the density plots in Fig.\ref{Figure8}. The linear potential plays an interesting roll as per as the energy confinement of the main lobe is concerned.
The growth rate of the peak power of the main lobe of a FEAP is found to be a function of potential strength $f$.  We illustrate this in Fig.\ref{Figure9}(a), where the peak power of the FEAP is plotted as a function of $\xi$ for different $f$ with a fixed value of $\delta_3 (=0.08)$. It is worthy to note that the peak power of the primary lobe of the pulse enhances monotonically for $f=-1$ compare to the case  when $f=0$ or $f=+0.5$ and reaches to a maxima at focusing point $\xi=1/3\delta_3$. On the other hand for $f>0$ the peak power reduces sharply. The confinement of power is not solely dependent on $f$ but also controlled by $\delta_3$. In Fig. \ref{Figure9}(b) we plot $P_{max}$ as a function of $\xi$ for a fixed potential strength ($f=-1$). As expected, the location of the focusing point (where $P_{max}$ reaches to a maxima) shifts according to $\delta_3$ value. It is also noted that the maximum power confinement is improved for relatively lower $\delta_3$. 

In order to understand the role of $f$ in energy confinement, in Fig.\ref{Figure9}(c) we plot peak power as a function of $f$ for a given distance $L  (L\leqslant \xi_{flip})$ using $\delta_3$ as a parameter. It is evident form this plot(Fig.\ref{Figure9}(c)) that the peak power reaches to a maxima for a specific potential strength (say $f_0$) which depends on $\delta_3$. The location of $f_0$ is indicated by the vertical dashed line which shifts towards more negative values for relatively lower $\delta_3$. If we analyse the solution derived in Eq.(\ref{eq4}) and minimize the decay factor we can get a compact analytical expression of  critical potential strength  $f_0=-{(3\delta_3 L)}^{-1}$, for which the peak power is maximized. Here $L  (L\leqslant \xi_{flip})$ is the  distance propagated by the pulse before reaching the flipping point. It is interesting to note that, at the limiting point ($L=\xi_{flip}$) the value of $f_0$ becomes $-1$ which is independent of $\delta_3$. In the following section we will try to understand why specifically at $f_0=-1$ we achieve maximum peak power. In order to validate the generalised expression, $f_0=-{(3\delta_3 L)}^{-1}$, for a given $L (< \xi_{flip})$ we calculate $f_0$ at which the peak power of the main lobe reaches to maxima and plot it as a function of $\delta_3$ in \ref{Figure9}(d) where numerical data (solid dots) agree well with  analytical formula (solid line).

\subsubsection{Region 2 ($\xi=\xi_{flip}$)}

Region-2 is the most important phase of Airy pulse propagation when $\delta_3>0$. In this region the propagating pulse faces a singularity and the unique characteristic of the Airy pulse is lost and it converges to a gaussian pulse. The area of the region-2 generally depends on the strength of TOD parameter $\delta_3$. For a very high $\delta_3$, the Airy pulse reaches to a finite-size focal area and region-2 can be squeezed to a point \cite{driben}. However, the tight focusing of the Airy pulse is not solely dependent on the strength of TOD but the linear potential may play a significant role.  Airy pulse loses its characteristic shape at singular point defined in region-2 and form a Gaussian pulse as,

\begin{equation}\label{eq6}
U(\xi_{flip},\tau)=U_0 \exp\left[\frac{-(\tau-\tau_1)^2}{\tau_f^2}\right]\exp(i\psi),
\end{equation}

where, the characteristic width is defined as $\tau_f=2\Delta/\sqrt{a}$ with $\Delta=\sqrt{a^2+\chi^2}$ and $\chi=(\xi_{flip}+3f\delta_3\xi_{flip}^2)/2 $. The detuned parameter $\tau_1=(a^2+\frac{f\xi_{flip}^2}{2}+\delta_3f^2\xi_{flip}^3$). The amplitude $U_0$ and phase $\psi$ are defined as
$U_0=\frac{1}{2\sqrt{\pi\Delta}}\exp(\frac{a^3}{3})$ and $\psi=\phi_0(\xi_{flip})-\frac{1}{2}\tan^{-1}(\frac{\chi}{a})+\frac{\chi(\tau-\tau_1)^2}{4\Delta^2}$, respectively. From the expression of the Gaussian pulse, that emerges at singular point, it is evident that the potential strength $f$ can control the area of region-2 by manipulating the characteristic width of the pulse. The dynamics of the Airy pulse for different $f$ are already shown in Fig.\ref{Figure7}. The figure indicates that the flipping area increases for positive $f$ value (Fig.\ref{Figure7}(a)) and  decreases for negative $f$ (Fig.\ref{Figure7}(d)). The flipping area merges to a focusing point (Fig.\ref{Figure7}(c)) for $f=-1$ where all  energy is confined. This feature is  unique and can be explained easily from the expression of the width of the Gaussian pulse that is formed at singular point $\xi_{flip}$. The full-width at half maxima (FWHM) of the pulse is defined as, 
 \begin{equation}\label{eq7}
 \tau_{FWHM}=2\sqrt{2\ln2}\left(a+\frac{(1+f)^2}{36a\delta_3^2} \right)^{1/2}.
 \end{equation}

  \begin{figure}[h!]
      \begin{center}
      \epsfig{file=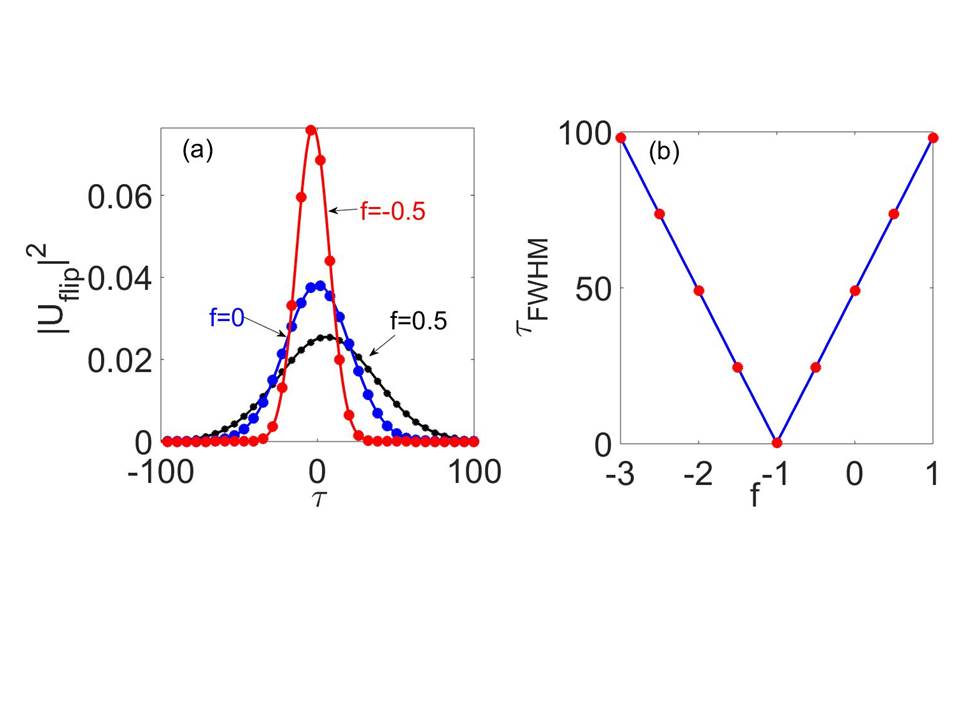,trim=0.2in 2.2in 0.1in 1.2in,clip=true, width=85mm}
      \vspace{0em}
     \caption{(a) The comparison of different  gaussian pulse obtained at $\xi_{flip}$ for different $f$. (b) The variation of the width of the gaussian pulse at the singular point with $f$ for $\delta_3=0.08$.Solid blue line represents the analytical result when red dots signifies the numerical data points.   } 
     \label{Figure10}
     \end{center}
     \end{figure} 

The area of the flipping region  depends on the width of the pulse.  Note, for a very large $\delta_3$ the $\tau_{FWHM}$ becomes smaller and the flipping region reduces to a point. Interestingly, the strength of the linear potential $f$ now acts as an additional variable to control the $\tau_{FWHM}$. In fact for $f=-1$ we have the minimum width and that is indeed the condition of absolute focusing (see Fig. \ref{Figure7}(c)).  In fig.\ref{Figure10}(a) we plot the Gaussian distribution of the Airy pulse at singular point for different values of $f$ and the variation of the width is evident. Generally the truncation parameter ($a$) is small and from Eq.\ref{eq7} it is easy to show that, at fixed $\delta_3$, $\tau_{FWHM}$ changes linearly with $f$ as $\tau_{FWHM}\approx |(1+f)|\xi_{flip}\sqrt{2\ln2/a}$ and becomes almost zero for $f=-1$. In Fig.\ref{Figure10}(b) we illustrate this feature where the width almost vanishes at $f=-1$ which corresponds to the absolute focusing point. This is consistent with the previous investigation where we find for $f=-1$ the primary lobe of the FEAP carriers maximum power if the pulse propagates up to the flipping point ($\xi=\xi_{flip}$).

\subsubsection{Region 3 ($\xi>\xi_{flip}$)}
    
    In this section we investigate the behavior of the FEAP beyond the flipping region ($\xi>\xi_{flip}$). Due to the perturbation of the TOD, the initial Airy pulse flips in temporal domain and propagates with a reverse acceleration \cite{driben,Roy}. The Airy pulse is converted to a Gaussian pulse at singular point ($\xi=\xi_{flip}$) and again forms an Airy pulse with inverted temporal wings for $\xi>\xi_{flip}$. Assuming the initial shape as Gaussian and resealing the distance parameter $\xi'=(\xi-\xi_{flip})$, it is possible to obtain the analytical expression of the inverted Airy pulse for $\xi>\xi_{flip}$ \cite{Roy}.

\begin{figure}[h!]
      \begin{center}
      \epsfig{file=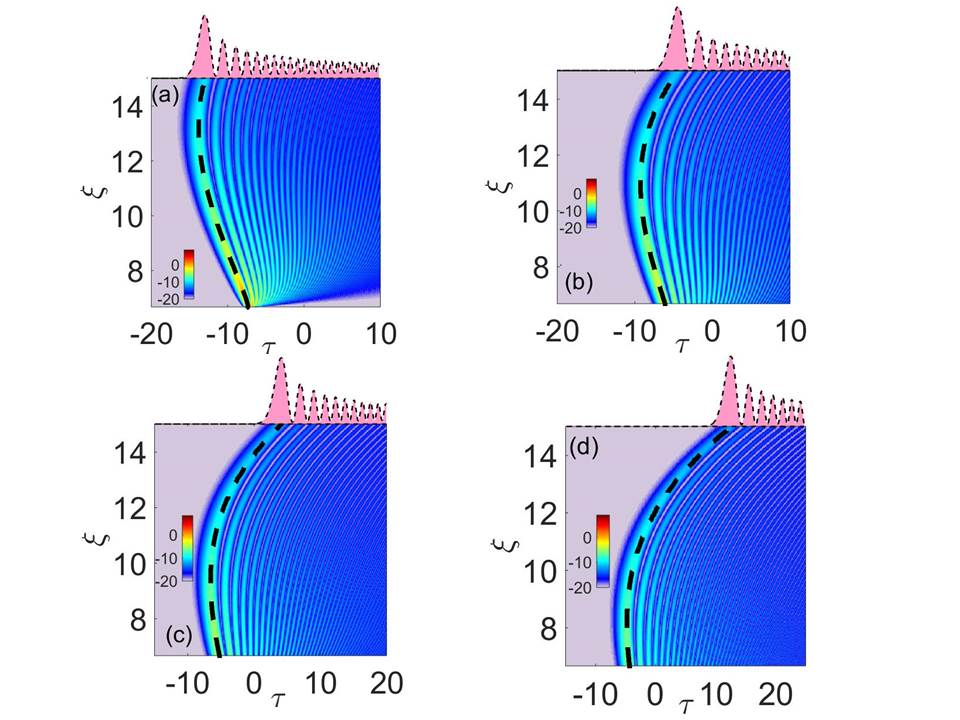,trim=0.8in 0.0in 1.0in 0.0in,clip=true, width=90mm}   
    \vspace{0em}
     \caption{The temporal dynamics of the pulse for different TOD parameter ($(a)\delta_3=0.05 (b) \delta_3=0.06 (c)\delta_3=0.07 (d)\delta_3=0.08) $ with $f=-1$. The comparison of analytical and numerical solution is shown in the upper panels of the figures. The black dashed lines represent the trajectory of the primary lobe which are superimposed with the numerical density plots. } 
     \label{Figure11}
     \end{center}
     \end{figure} 

However the exact solution for the inverted Airy pulse is not easy to obtain when $f\neq0$. The span of the flipping region (region 2 in Fig.\ref{Figure7}) is finite for $f\neq -1$. The region focuses to a point only when $f=-1$. Note that, the shape of the pulse is defined only at the flipping point ($\xi=\xi_{flip}$) and analytically undermined at other points in region 2. The knowledge of the analytical form of the pulse at flipping region is extremely essential since we use this shape as input to determine the inverted Airy pulse at output. So we determine the solution of the inverted Airy pulse at output for $f=-1$ as,

    \begin{equation}\label{eq8}
    u(\xi',\tau)=\frac{1}{c'}\exp(\frac{a^3}{3})Ai(\frac{b'}{c'}-\frac{n'^2}{c'^4})\exp i\left(\frac{2n'^3}{3c'^6}-\frac{n'b'}{c'^3}+\Phi\right)
    \end{equation}
where, $c'=(3\delta_3\xi')^\frac{1}{3}$ , $b'=-\tau+\tau_1-\frac{1}{2}f\xi'^2+\delta_3f^2\xi'^3$ and $n'=ia-\chi-\frac{1}{2}\xi'+\frac{3}{2}\delta_3f\xi'^2$. The total phase accumulated throughout the propagation is, $\Phi  =\frac{\delta_3f^3 }{4}(\xi'^4-\xi_{flip}^4)-\frac{1}{6}f^2(\xi_{flip}^3+\xi'^3)-f\xi'\tau$. In Fig.\ref{Figure11} we  plot the dynamics of the inverted Airy pulse at $\xi>\xi_{flip}$ for different $\delta_3$ parameter with $f=-1$. The analytical solution obtained in Eq.\ref{eq8} is plotted in the upper panels of the figure (dashed lines) along with its numerical counterpart (shaded area) and they agree well. The trajectory of the main lobe of the Airy pulse depends on the TOD coefficient ($\delta_3$)  and potential strength ($f$) as,

  \begin{equation}\label{eq9}
  \tau=\tau_1-\frac{f\xi'^2}{2}+\delta_3f^2\xi'^3-\frac{\Delta'^2}{c'^3},
  \end{equation}
      
 where $\Delta'=\chi+\frac{1}{2}\xi'-\frac{3}{2}f\xi'^2$. In Fig. (\ref{Figure11}) we plot the trajectory (black dashed line) obtained analytically through  Eq.\eqref{eq9}.  The analytical path corroborates well with the numerical trajectory of the main lobe.

\section {Conclusion}
In this paper we investigate the dynamics of a finite energy Airy pulse (FEAP) inside a Si-based waveguide where free-carriers are generated via a strong CW pump. The induced free-carriers lead to a dynamic change of refractive index in the time domain which acts like a liner optical potential experienced by the propagating FEAP. The usual parabolic trajectory of the FEAP is significantly influenced by the linear potential and it experiences a monotonous frequency shift. We theoretically solve the linear propagation equation with potential term and obtain a general solution of the FEAP under third order dispersion (TOD). We have separately investigated the effects of positive and negative values of third order dispersion on Airy trajectory. The dynamics of Airy pulse becomes robust under negative TOD and efficient manipulation of the trajectory is possible through linear potential. We derive the analytical expression of the pulse trajectory which agrees well with the numerical simulation. In case of positive TOD, the dynamics of the FEAP is dramatic as it experiences a singularity at a specific distance. The FEAP temporally flips at the singular point and then moves with a reverse acceleration. The entire dynamics of the FEAP is divided into three distinct regions and all the regions are studied thoroughly.    The linear potential plays a dominant role in energy confinement of the main lobe during the propagation of FEAP. This phenomenon has potential applications in remote energy transfer and pulse reshaping.  Nano-particle manipulation, optical routing are other few applications where manipulated Airy dynamics (through optical potential) may be useful.       
\section*{Acknowledgements}
A.B. acknowledges Ministry of Human Resource Development (MHRD), India for a research fellowship.

\end{document}